\documentclass{iau} 
\usepackage{amssymb}
\usepackage{graphicx}
\usepackage{caption}
\usepackage{url}

\def\HII{H\,\textsc{ii}}
\def\ha{H$\alpha$}

\def\HII{H\,\textsc{ii}}
\def\NII{[N\,\textsc{ii}]}

\def\OIII{[O\,\textsc{iii}]}
\def\SIII{[S\,\textsc{iii}]}

\title[Current planetary nebula surveys] 
{Challenges faced by current Galactic planetary nebula surveys}

\author[David Frew]   
{David J. Frew$^{1,2}$ }

\affiliation{$^1$The University of Hong Kong, Department of Physics, Hong Kong SAR, China\\ [\affilskip]
$^2$Laboratory for Space Research, The University of Hong Kong, Hong Kong SAR, China \\
email: {\tt djfrew@hku.hk}
}

\pubyear{2017}
\volume{323}   
\jname{PNe: Multi-wavelength Probes of Stellar and Galactic Evolution }
\editors{Xiaowei Liu, Letizia Stanghellini, \& Amanda Karakas, eds.}

\begin{document}

\maketitle

\begin{abstract}
Determining the demographics of the Galactic planetary nebula (PN) population is an important goal to further our understanding of this intriguing phase of stellar evolution. The Galactic population has more than doubled in number over the last 15 years, particularly from narrowband \ha\ surveys along the plane. In this review I will summarise these results, with emphasis on the time interval since the last IAU Symposium.  These primarily optical surveys are not without their limitations and new surveys for PNe in the infrared similarly face a number of challenges.  I will discuss the need for multi-wavelength approaches to discovery and analysis.  The desire to have accurate volume-limited samples of Galactic PNe at our disposal is emphasised, which will be impacted with new data from the Gaia satellite mission. We need robust surveys of PNe and their central stars, especially volume-limited surveys, in order  to clarify and quantify their evolutionary pathways.
 
\end{abstract}

\firstsection 
\section{Introduction}\label{intro}

Planetary nebulae (PNe) are a common end-point of stellar evolution and are an important contributor to the evolution of the Galaxy, but there is still much to learn regarding their mass-loss and nucleosynthetic processes and the importance of binarity in their formation.  Widely-held assumptions on their progenitor stars and visibility lifetimes have needed revision (Miller Bertolami 2016).   Ultimately, determining the ages, masses, and chemical compositions of a complete census of PNe is necessary to understand how low- to intermediate-mass stars influence the evolution of our Galaxy.
While close binary central stars have now been proven to have a shaping role for their PNe (Hillwig et al. 2016a), the common-envelope phenomenon is still poorly understood (De Marco, these proceedings).  Moreover, a subset of close binaries are double-degenerate systems, and potential SN\,Ia progenitors, which are important cosmological probes.

Many new Galactic PNe have been found over the last two decades, especially from narrow-band surveys along the Galactic plane, although these optical surveys are not without their limitations and undertaking new surveys for PNe similarly face a number of challenges.   In this review I will summarise these results, especially over the time interval since the last IAU Symposium. Here I will only concentrate on Milky Way surveys for PNe; for studies of their progenitors, the post-AGB stars and pre-PNe (e.g. Szczerba et al. 2012; Vickers et al. 2015) the reader is referred to   Lagadec (these proceedings).

\section{PN Surveys in the Milky Way}\label{surveys}
\begin{figure*}[h]
\centering 
\includegraphics[width=13cm]{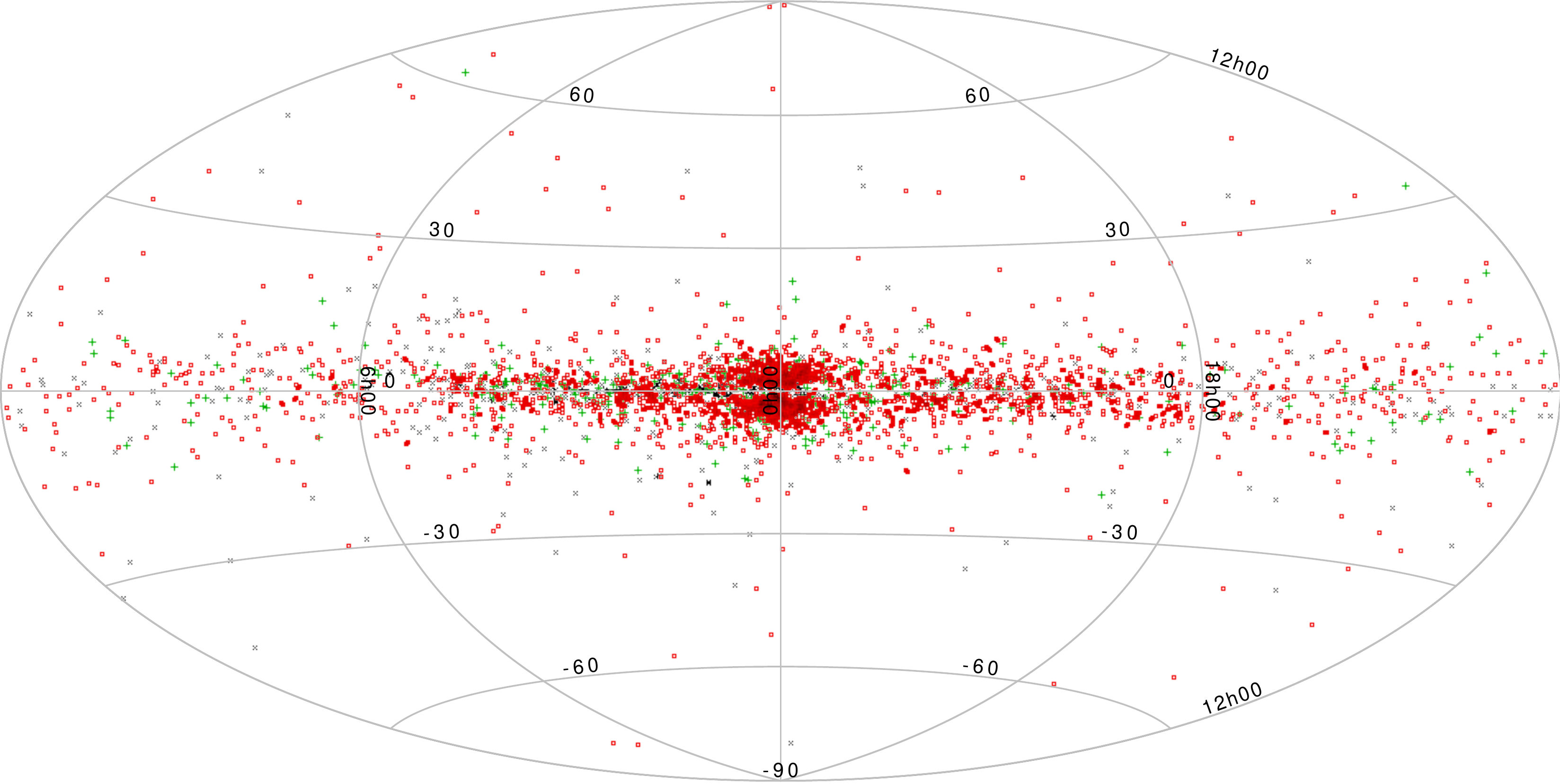}
 \vspace*{-0.1 cm}
\caption{Aitoff projection showing the Galactic distribution of all 3540 PNe currently in HASH. True, likely, and possible PNe are the red, green, and black symbols respectively.
}
\label{fig:Aitoff}
\end{figure*}

Current PN search techniques are imperfect in one way or another, with limitations depending on the spatial resolution and areal coverage of the survey, and on the size, surface brightness and extinction of individual PNe.  Using a range of complementary techniques should in principle recover most of the detectable PNe in a particular wavelength regime, assuming that each survey was systematically completed and covered the whole sky, which is clearly not the case.  Since the Spitzer surveys only covered a region very close to the Galactic plane (Benjamin et al. 2003; Carey et al. 2009), mid-IR surveys for PNe are quite incomplete,  Consequently, only a fraction of the true population of Galactic PNe has been recovered.  Past compilations have been seriously incomplete and biased, and only now has an attempt been made to consolidate all extant data sets into one database in a systematic manner (Parker et al. 2016a,b; Boji\v{c}i\'c et al. 2016). Currently, the Hong Kong/AAO/Strasbourg/\ha\ (HASH) database contains 3540 true, likely, and possible PNe (Parker et al. 2016b). This number is only about 15--30\% of the estimated total of Galactic PNe (Frew 2008; Jacoby et al. 2010), showing that only a small fraction of Galactic PNe have been catalogued.  The distribution of the current population of Galactic PNe is seen in Fig.\,\ref{fig:Aitoff}.

Most of these undiscovered PNe are invisible at optical wavelengths and future confirmations and discoveries will require surveys in the IR and radio domains across wide swaths of the sky with enhanced sensitivity and resolution. Since the work of Boji\v{c}i\'c et al. (2011), some progress has already been made in these areas (e.g.  Hoare et al. 2012; Norris et al. 2011; Umana et al. 2015).  Ultimately, IR imaging and spectroscopy will be required to validate future discoveries in a similar way to current classification protocols.

\section{Detection Techniques, Past and Present}\label{techniques}
\begin{figure*}
\centering  
\includegraphics[width=10cm]{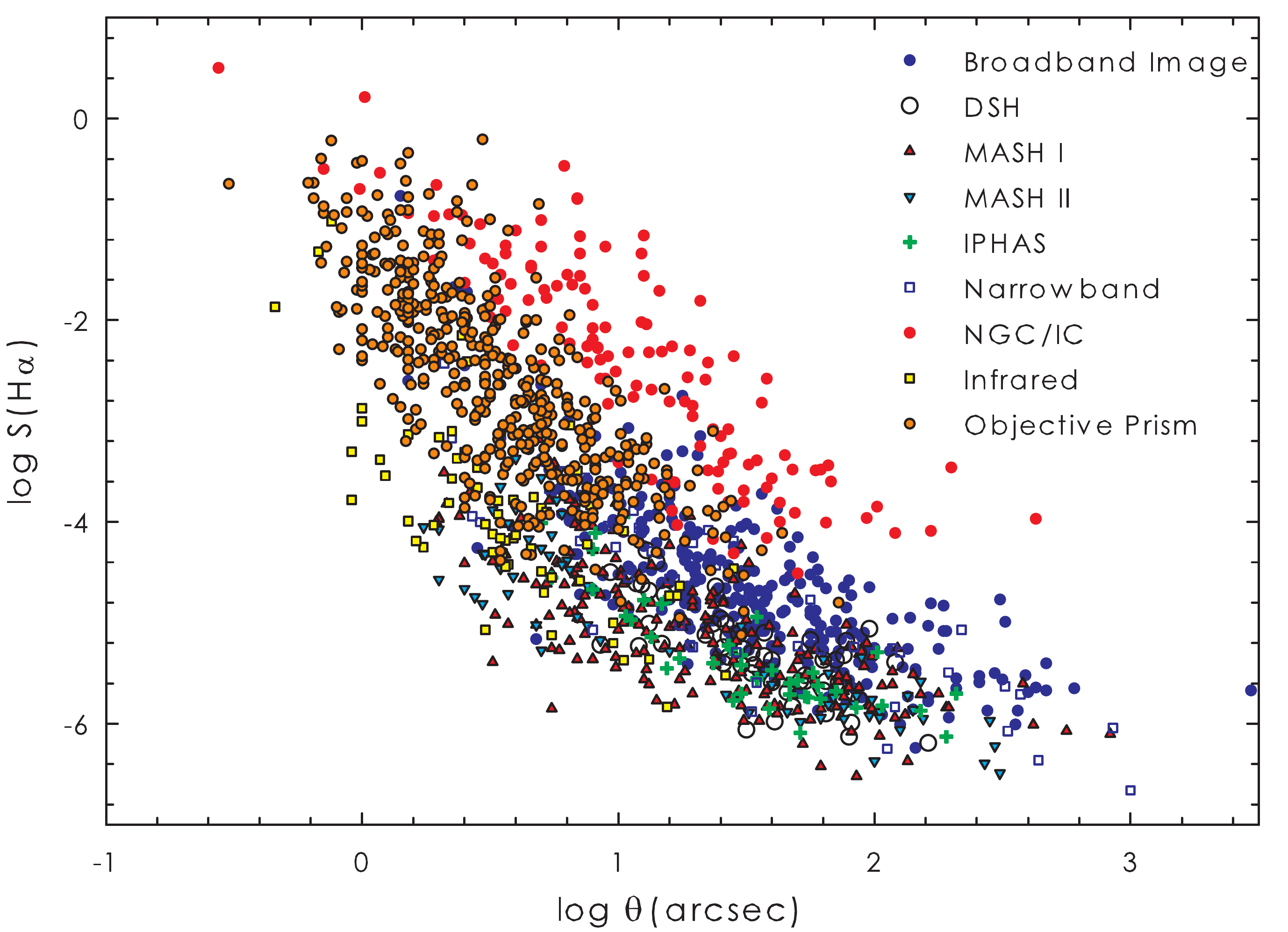}
 \vspace*{-0.1 cm}
\caption{PNe with reliable \ha\ surface brightnesses from HASH. Here the surface brightness is plotted against nebular radius.  Different discovery techniques are sensitive to different samples of PNe, as outlined in \S\ref{techniques}.
}
\label{fig:S_theta}
\end{figure*}

It is germane to briefly review the various detection techniques that have been used to discover Galactic PNe over the last two centuries (for a detailed review, see Frew \& Parker 2010).  Each technique is sensitive to PNe of differing size and brightness as illustrated in Fig.\,\ref{fig:S_theta}, and summarised below:

\begin{itemize}
\item Visual discoveries, typified by the NGC PNe (e.g. those found between 1780 and 1838 by W. and J. Herschel), which are relatively large \textit{and} bright (Fig.\,2).

\item Broadband photographic imagery e.g. Abell (1966), and the recent work by the Deep-Sky Hunters (DSH) team, an amateur-professional consortium (Jacoby et al. 2010; Kronberger et al. 2016, and references therein).  Candidates were typically identified from direct inspection of images, or the online scanned data in the case of the DSH discoveries.

\item Narrowband imagery, e.g. the MASH catalogues (Parker et al. 2006, Miszalski et al. 2008), IPHAS catalogue (Sabin et al. 2014), and some finds by the DSH team (Kronberger et al. 2014, 2016), and other smaller surveys (Fig.\,2); and potentially with the J-PAS survey (Gon\c{c}alves et al. 2016). These PNe are usually of low surface brightness, being identified via direct inspection, image processing, or using color-color plots.  

\item Near-infrared (near-IR) imagery, particularly utilising narrowband images to find faint PNe;  e.g. Jacoby \& Van de Steene (2004) in the \SIII\ line, and discoveries from the UWISH2 survey (Froebrich et al. 2015; Gledhill, these proceedings) in the light of molecular hydrogen.  Near-IR imaging of IRAS mid-IR candidates has also been useful for classifying new PNe and pre-PNe (Ramos-Larios et al. 2009, 2012)

\item Mid-IR imagery or colors, with radio or near-IR confirmation in some cases (e.g. Preite-Martinez 1988; Van de Steene \& Pottasch 1993; García-Lario et al. 1997; Kwok et al. 2008; Cohen et al. 2010; Mizuno et al. 2010; Ingallinera et al. 2016).  WISE mid-IR imagery has also been used to find very faint PNe (Kronberger et al. 2016).

\item Low-resolution objective-prism spectroscopic surveys by many authors, particularly before the 1970s, that revealed modestly bright but generally compact PNe.

\item Higher-resolution spectroscopic discoveries using multiplexed fiber spectroscopy, e.g. Yuan \& Liu (2013), and \S\ref{serendip}.

\item Direct discoveries of low-mass [WR] central stars (Kanarek et al. 2015, 2016), that may be surrounded by currently undetected faint PNe. This approach has the potential to find many obscured H-deficient central stars.
\end{itemize}

\smallskip
Most discoveries in the last decade have utilised narrowband imaging or near- and mid-IR surveys.  Since the last IAUS review in 2011 (Parker et al. 2012), over 750 PNe and PN candidates have been found, many still needing assessment in the HASH database.  Sabin et al. (2014) announced the discovery of 159 true, likely and possible PNe, 284 PN candidates have been found from UWISH2 (Froebrich et al. 2015; Gledhill, these proceedings), and another 180 from the DSH team and related efforts (Kronberger et al. 2016; and personal communication; Acker et al. 2015). Fragkou et al. (2016) have found 70 candidates using radio and IR surveys, and other contributions in the infrared have been made by Flagey et al. (2014), Nowak et al. (2014), and Ingallinera et al. (2014a,b, 2016) who are systematically investigating the 300+ unidentified MIPSGAL nebulae found by Mizuno et al. (2010), though most remain un-assessed at the time of writing.

Small additional lists of new objects or one-off discoveries, not including our own unpublished objects, include those of Ramos-Larios et al. (2012), Parker et al. (2012b), Miszalski et al. (2013), Aller et al. (2013), Blanco C\'ardenas et al. (2013), Frew et al. (2013), Munari et al. (2013, see Frew et al. 2014c), Yuan \& Liu (2013), Miranda et al. (2014), Miszalski \& Miko\l ajewska (2014), Chhetri et al. (2015) and Kanarek et al. (2016).

\section{The Problem of PN Mimics}\label{mimics}
Contaminating objects in previous PN surveys have been the rule, and may never be reduced to zero, though Parker et al. (2016a,b) has explained the rationale for HASH. Both emission-line stars and extended objects of various types have been confused with PNe (for a review, see Frew \& Parker 2010).  Emission-line stars, historically found through objective prism surveys, and now from multi-band CCD surveys, can be subdivided into several classes: symbiotic stars, Be and B[e] stars, Wolf-Rayet stars, luminous blue variables (LBVs), pre-main sequence stars, cataclysmic variables of various types, and even late-type giant stars, with or without Balmer emission.

Extended nebulae, usually found via imagery, have as mimics included \HII\ regions, reflection nebulae, galaxies, symbiotic outflows, Wolf-Rayet shells and LBV ejecta nebulae, supernova remnants, young stellar objects, Herbig-Haro objects, other miscellaneous nebulae and even plate defects.  Indeed, several of the nearest candidates are likely to be Str\"omgren zones in the interstellar medium ionized by a hot pre-white dwarf or subdwarf, with any original PN having long since dispersed.   If the ionizing star is hot, these \HII\ regions can have \OIII\ emission, and emission-line ratios which mimic true PNe; e.g. Hewett\,1, DeHt\,5,  Sh\,2-68, and Sh\,2-174 (Hewett et al. 2003; Chu et al. 2004; Frew 2008; cf. Ransom et al. 2010, 2015).

A suite of diagnostic diagrams are now available to help eliminate these mimics and to discriminate between them and bona fide PNe, transitional objects between the post-AGB and PN phases, and a number of peculiar PN-like nebulae (Anderson et al. 2010; \cite[Frew \& Parker 2010]{FP10}; Sabin et al. (2013); Frew et al. (2014b, 2016a,b).  Symbiotic stars and their resolved nebulae, if present, remain a thorn in the side of classification efforts. Corradi et al. (2010) have devised useful diagnostic plots for symbiotic stars, to complement the diagrams mentioned previously.

\section{Volume-limited Surveys}\label{census}
Extant PN surveys in the Milky Way, and the studies utilising them, have until now been flux-limited, and biased against the most abundant kind of PN -- evolved ones!  It should be apparent that volume-limited surveys are needed for unbiased statistical analysis of PNe and their central stars (multiplicity, central star properties, nebular properties, nebular kinematics), and for robust comparison with population synthesis models and evolutionary theory.  Specifically we need to clarify and quantify their evolutionary pathways, which may be several in number (Frew \& Parker 2010; Reindl et al. 2014; Hillwig et al. 2016).  
Furthermore the discovery of Galactic PNe continues, including several nearby objects, discussed in \S\ref{techniques} and \S\ref{serendip}.

Current developments on volume-limited surveys are being undertaken by Frew et al. (in prep.).  At present, relatively few accurate primary distances are available (see \S\ref{DR1}), so we have relied on statistical distances; e.g. Stanghellini et al. (2008), and most recently, Frew et al. (2016a).  To date, only a few studies have utilised volume-limited samples, but particularly noteworthy are the ChanPLaNS (Kastner et al. 2012; Freeman et al. 2014; Montez et al. 2015) and HerPlaNS (e.g. Ueta et al.  2014; Aleman et al. 2014) surveys in the X-ray and thermal-IR domains respectively.

\section{A short look at Gaia Data Release 1}\label{DR1}
The local census of PNe will be greatly helped in future with data from the Gaia survey (Gaia Collaboration 2016).
Gaia Data Release 1 (Lindegren et al. 2016) contains positions and $G$ magnitudes (in the Gaia system) for some hundreds of central stars, but less than two dozen have  proper motions and parallaxes (Stanghellini et al. 2016; Manteiga, these proceedings), and only four of these parallaxes are statistically meaningful, with errors of $<$50\%).  To make a better comparison with the latest statistical distance scales, the Gaia data are supplemented with the HST fine-guidance sensor parallaxes (Benedict et al. 2009), and high-quality USNO data from Harris et al. (2007).  A comparison of the trigonometric distances against the statistical distances from Frew et al. (2016a) show good agreement (Fig.~3), but less so with the older Stanghellini et al. (2008) distance scale (see Smith 2015).  Two strongly discrepant points, Cn~3-1 and M~1-77, are both young, compact objects (not shown in Fig.~3). Their Gaia distances are only 520 and 930\,pc respectively, compared to estimates of 4 -- 6\,kpc from both statistical scales.  Either the preliminary Gaia parallaxes for these two compact nebulae are seriously in error, or we are dealing with a new PN-like phenomenon at low luminosities.

\begin{figure*}[h]
\centering 
\includegraphics[width=6.63cm]{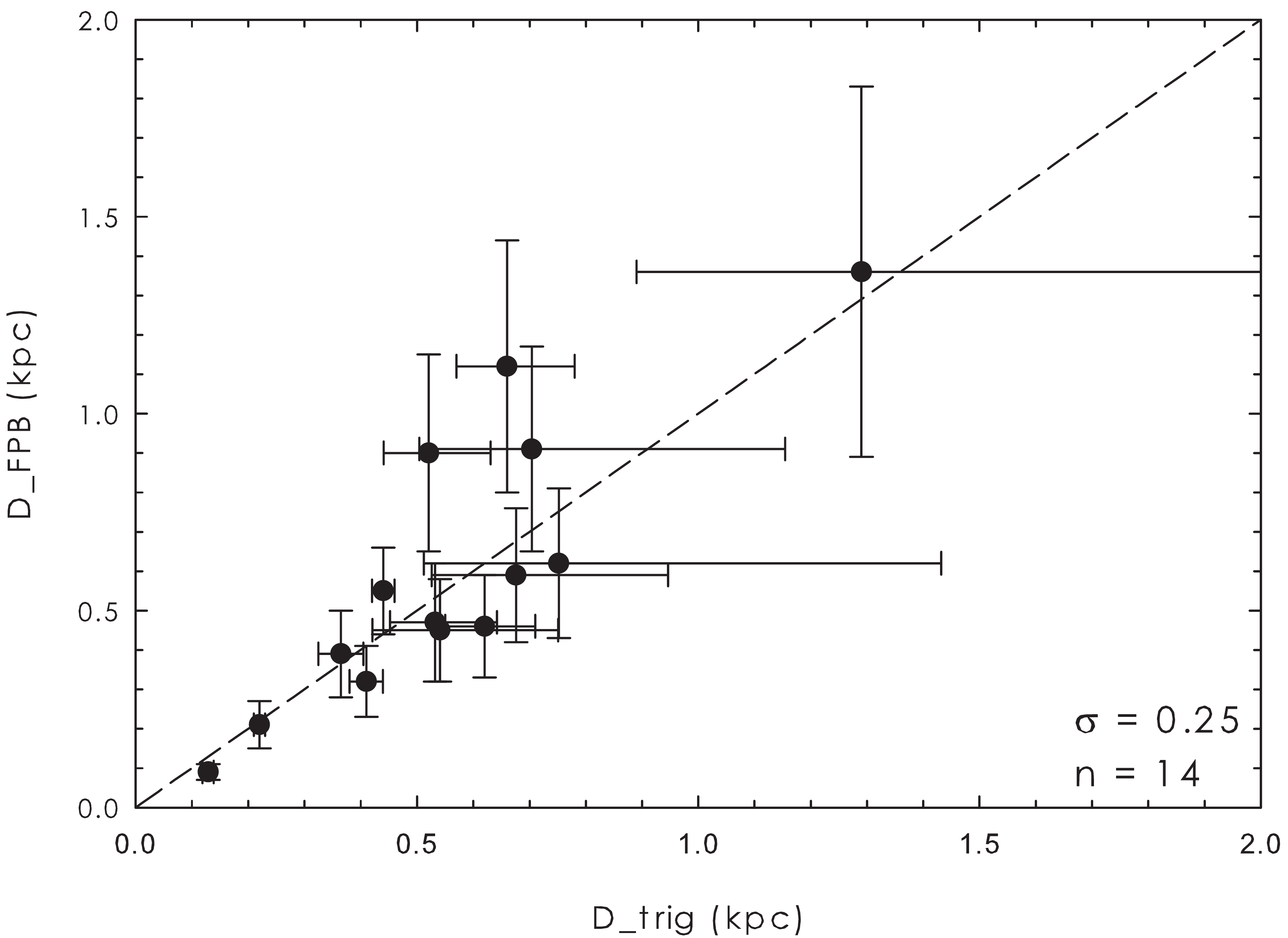}
\includegraphics[width=6.63cm]{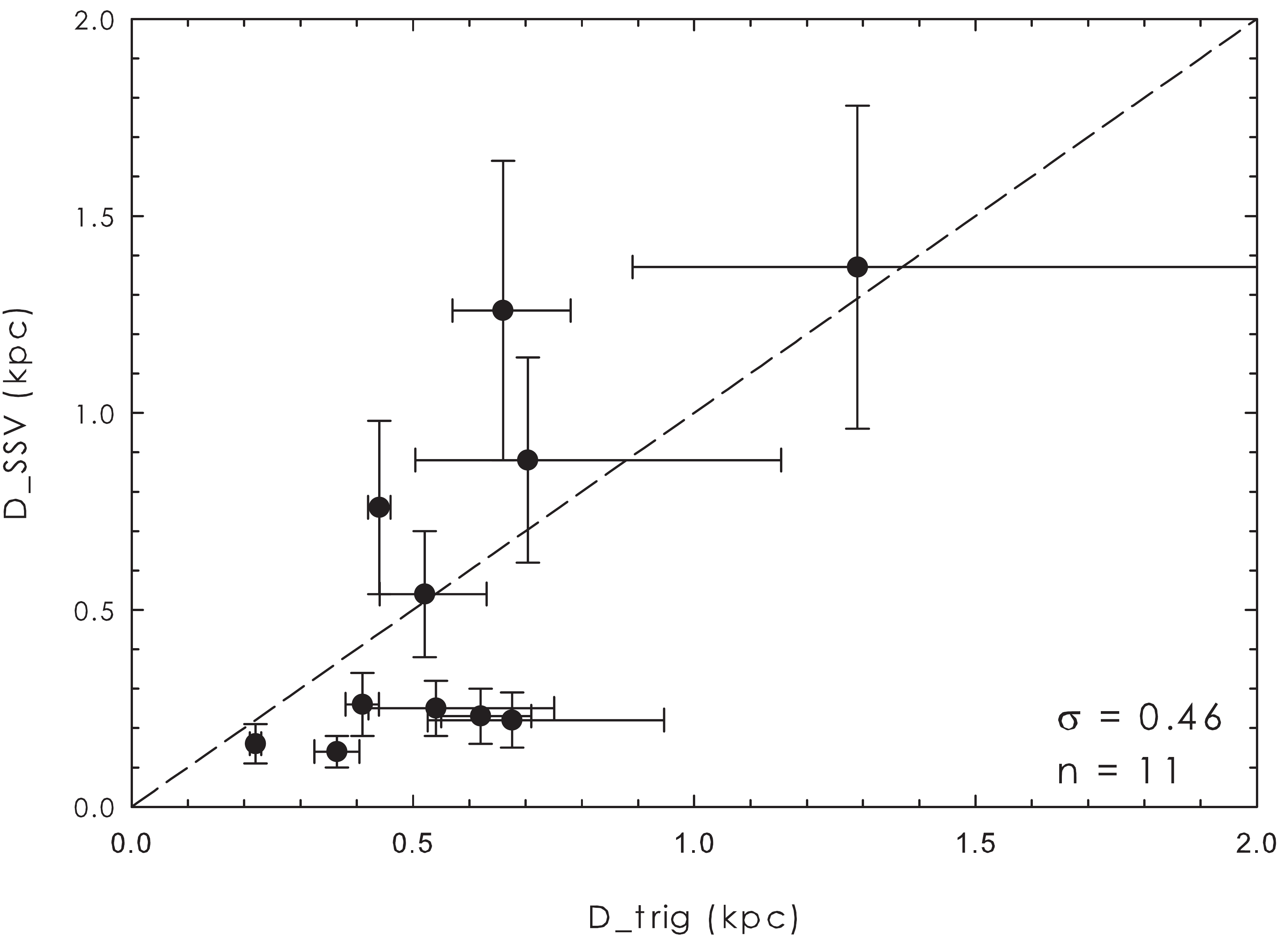}
\caption{Comparison of trigonometric distances with published statistical distances. The comparison is with the Frew et al. (FPB) scale (left) and the Stanghellini et al. (SSV) scale (right).
}
\label{fig:trig}
\end{figure*}

Even after later Gaia data releases become available, statistical distances will remain complementary.  In the optical, the \ha\ emission-line is the most useful calibrating waveband, as it is brighter than H$\beta$, is less affected by moderate reddening, and because several flux-calibrated  \ha\ imaging surveys are now available (\cite[Gaustad et al. 2000]{SHASSA}; Drew et al. 2005, 2014; \cite[Frew et al. 2014a]{F14a}), from which large numbers of accurate integrated \ha\ fluxes can be determined (Frew et al. 2013, 2014a).  For highly-reddened nebulae, Paschen-$\alpha$ is the brightest detectable hydrogen line but is only accessible from space (Wang et al. 2010), but Brackett-$\gamma$ is also useful in the near-IR.  Forbidden mid-IR fine-structure lines  are almost immune to reddening, but the intensities are strongly influenced by the excitation of the PN, and thus are more problematic to calibrate.  Ultimately, the advent of accurate radio-continuum fluxes for the many small, faint PNe current lacking in data (e.g. Norris et al. 2011) will warrant a recalibration of the radio distance scale.

\section{The View from Serendip}\label{serendip}
As I was finishing this review, an interesting paper by N\'u\~nez et al. (2016) appeared in preprint form, on chromospheric emission in late-type members of the rich open cluster M\,37 (NGC 2099).  Fiber spectra of some of the stars in the northern part of the cluster showed contaminating forbidden \NII\ emission lines which were strong relative to \ha, typical of an evolved PN.  The authors found a nebula on IPHAS \ha\ imagery but made no firm conclusion on its nature.  An evening spent with online databases revealed a 19-mag blue star near the centre of the box-shaped nebula on the DSS blue image.  Applying the well-established cluster reddening to the available GALEX/SDSS photometry gives colors consistent with a hot star, and if a member of M\,37 ($D$ = 1.6\,kpc), the absolute magnitude ($M_{\rm v}$ = +7.5) is typical of an evolved ionizing star.  Curiously, this star was already noted as a candidate white dwarf by Gentile Fusillo et al. (2015) and independently found by Chang et al. (2015) to be a periodic variable with a period of 0.445 days.  This is undoubtedly a new bipolar PN, and potentially a physical member of M\,37, so is of great interest, especially if the CSPN is a close binary (see Jones 2016).  We are planning to follow-up on this object shortly. This serendipitous discovery shows that our census of even relatively nearby PNe is still incomplete.

\section{Future Directions}
Future discovery surveys in the Milky Way will continue to be useful, as part of efforts to characterise the entire Galactic PN population (Parker et al. 2016a,b). More powerful imaging and spectroscopic IR surveys of representative samples of PNe are necessary to better understand mass-loss processes and dust chemistry.  The advent of the Large Synoptic Survey	
Telescope (LSST) will be a major breakthrough in time-domain studies of central stars in particular. For the nebulae, high-resolution spatial and kinematic and spatial surveys with ALMA will be rewarding, for both discovery, and importantly in the characterisation of post-AGB stars and younger PNe.  The flagship James Webb Space Telescope (JWST), to be launched in 2018, will potentially have a major impact on Galactic PNe studies, though less so in the mid-IR owing to its faint saturation limits (Barlow 2012). The recently launched Astrosat mission should also provide useful complementary data at X-ray and UV wavelengths.
Continued work on volume-limited surveys will constrain more tightly the relative statistics of different PN populations, and this is where where the final Gaia data release at the end of the decade will have the greatest impact.  In addition, the common envelope phenomenon remains poorly understood, as is its relationship to the newly recognised abundance discrepancy problem (Corradi et al. 2015).  We need deep spectra of all central stars in the local volume (cf. Weidmann \& Gamen 2011) for comparison with theoretical models, and to also constrain the relative frequency of binary and ternary systems (see Akashi, these proceedings), and their potential role in forming and shaping  PNe.

\newpage


{}

\end{document}